\documentstyle[prb,aps,multicol,epsf]{revtex}

\def\addcontentsline#1#2#3{\relax}
\def\beqn{\begin{eqnarray}}
\def\eeqn{\end{eqnarray}}
\def\beqns{\begin{eqnarray*}}
\def\eeqns{\end{eqnarray*}}
\def\beq{\begin{equation}}
\def\eeq{\end{equation}}
\def\bea{\begin{array}}
\def\ea{\end{array}}
\def\one{1\hskip-.37em 1}
\def\<{\langle}
\def\>{\rangle}

\begin{document} 
%
\draft
\title{Anomalous electronic conductance in quasicrystals}
\author{Stephan Roche}
\address{${\ }^{*}$ Department of Applied Physics, University of Tokyo, 7-3-1
Hongo, Bunkyo-ku, Tokyo 113-8654, Japan.}
\date{\today}
\maketitle
\begin{abstract}
Contributions of quantum interference effects occuring in 
quasicrystals are reported. First conversely to 
 metallic systems, quasiperiodic ones are shown to enclose
 original alterations of their conductive properties 
 while downgrading long range order. Besides, origin
 of localization mechanism are outlined within the context 
 of the metal-insulator transition (MIT) found in these materials.
\end{abstract}

\multicols{2}

\section{Introduction}

Despite sustained effort and concern, today's understanding of exotic 
electronic properties of quasicrystals \cite{AlMn} remains unsatisfactory 
although quasicrytalline materials have already been implemented to 
miscellanous concrete applications\cite{QC-app,QC-app2}. In particular, the role of quasiperiodic order on 
electronic localization and transport is believed to genuinely entail the most 
unexpected experimental features whereas so far, no coherent theoretical 
framework has been really successfully ascertained\cite{Berger,RocheJMP,Takeo,RocheFS}. As 
a matter of fact, one of the unprecedented tendency of quasicrystals
 is the enhancement of their conductive ability upon increasing 
contribution of static (structural disorder) or dynamical excitations 
(phonons). This has been strongly supported by many experimental 
evidences\cite{Berger} and is often refered as an original 
property in the litterature. Notwithstanding theoretically, 
given heuristical arguments\cite{RocheJMP,Sire} and numerical 
investigations (e.g. the Landauer conductance for quasiperiodic Penrose lattices 
\cite{Fujiwara-Land} or Kubo formula for 3D-quasiperiodic models \cite{RochePRL1}
) yield to uncomplete understanding of the observed properties which 
range from anomalously metallic behaviors to insulating ones 
\cite{Julien}.

It is generally assumed that a specific ``geometrical localization process'' takes place in quasicrystals 
(sustained by critical states\cite{Kho,Macia}) and that local disruptions of 
corresponding mesoscopic order reduce quantum interferences, resulting 
in an increase of conductivity. This issue has been takled for 1D 
quasiperiodic potential, tight-binding (TBM) 
as well as continuous Kronig-Penney models, and phason-type disorder
 has been shown to disclose manifestations of quantum interferences 
 (QIE) in quasiperiodic order \cite{RocheFas}. 

Pionneer works of M. Kohmoto\cite{Kho} on 
multifractal properties of critical states in 1D-quasiperiodic chains 
have been recently followed by renewed focus on the relation between 
localization features of such states and their ability to convey 
current\cite{Macia}. Still, variety of critical states let the 
question about localisation properties and transport ability
relation unresolved. The effect of disorder on top of these states is 
thus a complicated issue and very scarce rigorous results are available in the 
litterature. Attempts to rigorously establish analytical results in 
high-dimensional quasiperiodic systems have been facing 
some limitations despite encouraging early attempts
\cite{Moulopoulos,Moulopoulos-2}.

In what follows, we review exact results carried out on 1D 
umperfected quasiperiodic systems. We also give additional and 
complementary results to enlarge the understanding of an early study\cite{RocheFas}
Besides, the role of quantum interferences on both sides of the
 quasicrystalline MIT for higher dimensional materials 
 is outlined in a second part. A general scenario to follow the 
 metal-insulator is drawn in the ligth of recent results.

\section{Intereference effects in 1D quasiperiodic systems with 
disorder}

Introducing disorder could be done through typical
randomizing of site or hopping energies, with the subsequent 
occurence of Anderson localization in the infinite chain limit. For 
finite systems, localization lengthes may be much larger than the 
characteristic size so that conductance fluctuations as a function of 
energy of tunneling electrons (from the leads to the system) keep its 
self-similar nature and still follow power law behavior for system 
size studied in \cite{Dasarma} even when introducing disorder of 
$10\%$ of the total bandwith. However, the particular order sustaining long 
range quasiperiodicity suggest the possible presence of unique defects, known 
as phason-type defects. Their geometrical definition and
 properties have been subjected to many 
studies\cite{Phasons}, although some aspects remain controversial. 
From certain viewpoint it seems natural to consider how disruptions of 
quasiperiodic order inherent to such systems will 
degrade or improve transport properties. It is the aim of this work 
to contribute to such more general and fundamental understanding of 
electronic propagation in quasicrystals.

For 1D-quasiperiodic systems, we can define phasons 
that keep the essential characteristic of real systems, in the 
sense that they are a generic form of disorder which has no equivalent in usual metallic 
and periodic systems. In a preliminary study, such phason defects
were introduced by K. Moulopoulos, as a main probe for investigating 
Landauer conductance\cite{RocheFas}.

Tight-binding models (TBM) of perfect quasiperiodic 
chains have been intensively worked out both analytically and numerically only for some
given energies, but the results are supposed to have provided typical features of 
localization in quasiperiodic structures, such as power-law decrease 
of wavefunctions or power-law bounded resistances \cite{Kho}. However, 
if leading to interesting analytical results, 
TBM do not allow to simply investigate energy-dependent properties of quantum dynamics
and electronic transport. Following the works and 
discussions of Kollar and S\"uto\cite{Suto}, and
 M. Baake et al.\cite{Baake}, the limitations of 
TBM when may be shown when considering effects of phason-type disorder 
on electronic localization and propagation. Let us recall the main 
results of \cite{RocheFas}. Tight-binding expression of the hamiltonian 
usually reads ${\cal H}=\sum_{n}t_{n}(|n\>\<n+1|+|n+1\>\<n|)$ 
($\gamma =t_{A}/t_{B}$ will stand for intensity of quasiperiodic 
potential, following a Fibonacci sequence ABAABABAABAAB\ldots, whereas site energies are kept constant)  and the 
Schr\"{o}dinger equation on a localized basis gives

{\small
\beq
\left(
\bea{c}
\psi_{n+1} \\
\psi_{n} \\
\ea
\right)=M_{n}
\left(
\bea{c}
\psi_{n} \\
 \psi_{n-1} \\
\ea
\right)=M_{n}M_{n-1}....M_{1}
\left(
\bea{c}
\psi_{1} \\
\psi_{0} \\
\ea
\right)=P_{n}
\left(
\bea{c}
\psi_{1} \\
\psi_{0} \\
\ea
\right)\nonumber
\eeq
}
with $\psi_{n}$ the component of wavefunction
for energy E at site n 
\beq
M_{n}=
\left(
\bea{cc}
0 &
-{\displaystyle {t_{n-1}}\over{\displaystyle t_{n}}}\\
{\displaystyle 1} & {\displaystyle 0}\\
\ea
\right)
\ \ \ \ \ \hbox{and} \ P(n)={\displaystyle \prod_{i=1}^{n}} M_{i}
\nonumber
\eeq

\begin{figure}[htbp]
\leavevmode
\epsfxsize=8cm
\centerline{\epsfbox{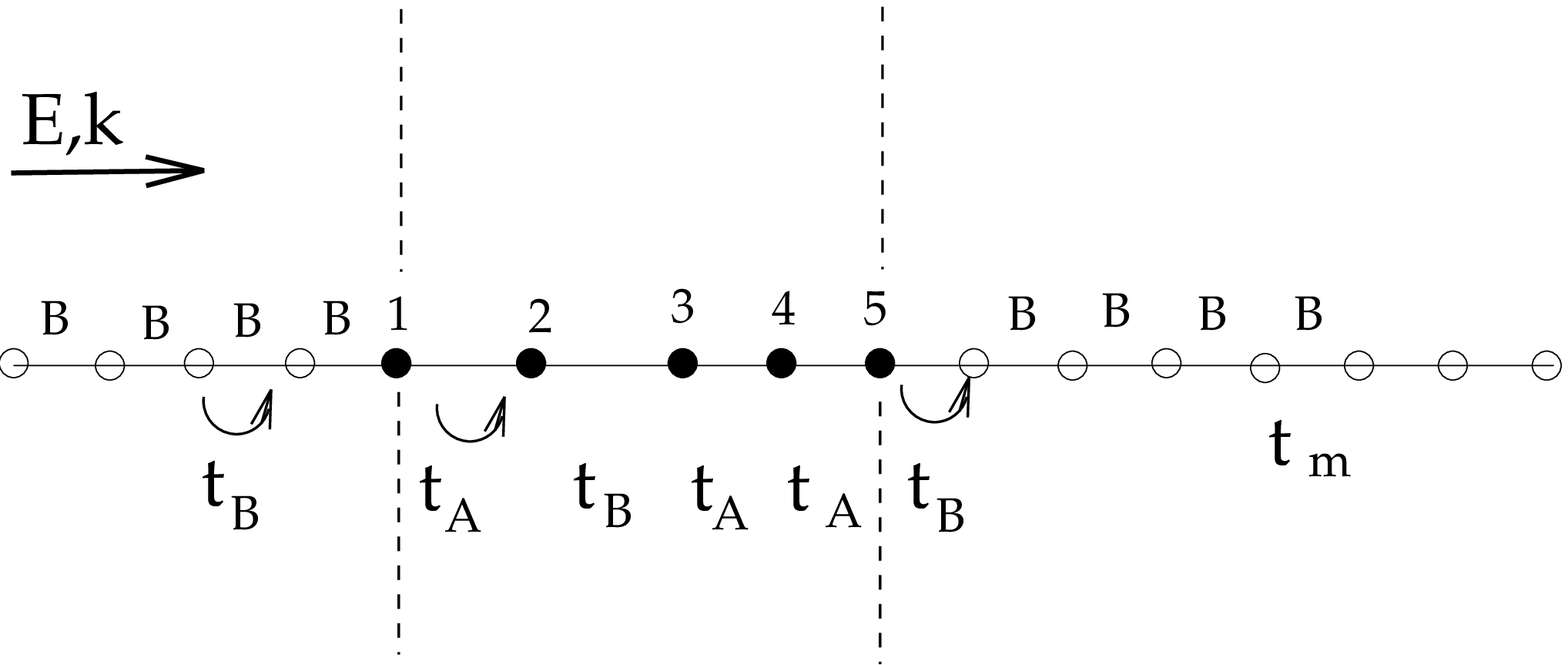}}
\narrowtext{\caption{Chain $N=4$: ABAA connected with perfect leads.}}
\label{figLF_s}
\end{figure}
Overlap integrals are given by $t_{n}=f(\theta_{n})$ 
with $\theta_{n}$ \cite{HKubo} is a kind of phase and $n$ a link 
index, so that $\theta_{n}=\frac{n}{\tau}+\theta_{0} (mod 1)$
\begin{eqnarray}
f(\theta)&=& t_{A} \ \ \ \hbox{for}\ \  \ 1/\tau^{2}\leq \theta < 1 
\nonumber \\
         &=& t_{B} \ \ \  \hbox{for}\ \  \ 0 \leq \theta <  
1/\tau^{2}  \nonumber
\end{eqnarray}
where $\theta_{0}=0$ for the usual Fibonacci chain. The quasiperiodic 
chain corresponding to an initial phase $\theta_{0}=m/\tau$ 
is associated with the Fibonacci sequence taken at 
$m^{\hbox{th}}$ site of the usual one. We define the phason as a 
abrupt geometrical transition between the two chains with 
$\theta_{0}=0,2/\tau$. The function $f(\theta_{n})$ 
will take two values $t_{A}$ or $t_{B}$, according to the link.
This gives a deterministic way to construct 
phason defects, which break long range quasiperiodic order, but 
which is a kind of disorder which has no equivalent in periodic 
systems. Introducing a so-called local isomorphism class by changing 
initial phase $\theta_{0}$ randomly chosen between 0 and 1, 
statistical studies of the localization and transport properties of 
Fibonacci chains have been performed\cite{HKubo,Wij}. 
\subsection{Effect of phason-defects at E=0 with TBM}

In a transmission study, we define the boundary 
conditions as $t_{N}=t_{0}=t_{ext}=t_{B}$ (Fig. 
\ref{figLF_s}), which lead to special values of the number of sites.
From the two Fibonacci chains with $\theta=0,2/\tau$, the corresponding
sequence is  $N(i=1,2,3,4....)=4,12,17,25,33,38,46,51,59,...$
noticing that the difference between two consecutive numbers follow a 
Fibonacci sequence with numbers 8 and 5 (8,5,8,8,5,8,5,8...). 
\endmulticols\widetext
\vspace{-6mm}\noindent\underline{\hspace{87mm}}
\begin{eqnarray}
\hbox{I} &:&(B)\hbox{{\bf 
ABAABABAABAABAB}}\hbox{AABABAABAABABAABAABABAA}(B) \ \theta_{0}=0 
\nonumber\\
\hbox{II} &:&(B)\hbox{AABABAABAABABAA}\hbox{{\bf 
BABAABAABABAABAABABAABA}}(B) \ \theta_{0}=2/\tau
 \nonumber\\
\hbox{III} &:&(B)\hbox{ABAABABAABAABA}{\bf 
BB}\hbox{ABAABAABABAABAABABAABA}(B)\nonumber
\end{eqnarray}
%
\noindent\hspace{92mm}\underline{\hspace{87mm}}\vspace{-1mm}
\multicols{2}
\noindent
In what follows, the N(i+1)-term of the sequence ${\cal S}$ will not 
correspond to N+1 sites but to the number of allowed sites as deduced 
from the above-mentionned Fibonacci sequence. Fig. \ref{figLF_s} shows a small chain with 
5 sites. The construction of a phason is illustrated below for a N=29 
site chain. Taking $\gamma =t_{A}/t_{B}$, transfer matrix can be evaluated analytically as
 well as the Landauer resistances. Defining the matrices 
 ${\cal A}, {\cal B}, {\cal C}$ by
\beq
{\cal A}=
\left(
\bea{cc}
0 &  -\frac{1}{\gamma} \\
\gamma & 0 \\
\ea
\right)
\ \ \ 
{\cal B}=\left(
\bea{cc}
\gamma & 0 \\
0 & \frac{1}{\gamma} \\
\ea
\right)
\ \ \ \ 
{\cal C}={\cal B}{\cal A}=
\left(
\bea{cc}
0 &  -1 \\
1 & 0 \\
\ea
\right)
\eeq
One shows that the sequence $P_{N}$ follows a Fibonacci sequence.
Besides, noticing that ${\cal CA}={\cal C}$, ${\cal AC}={\cal 
CA}^{-1}$ and ${\cal C}^{2}=-\one$, one shows that whatever $P_{N}$ 
there exist a recurrent structure given by ${\cal 
C}^{t(N)}{\cal A}^{s(N)}$
\beq
P_{N}=
\left(
\bea{cc}
0 &  -1 \\
1 & 0 \\
\ea
\right)^{t(N)}
\left(
\bea{cc}
\gamma &  0 \\
0& \frac{1}{\gamma}\\
\ea
\right)^{s(N)}
\eeq
where $t(N)$ are integers and  $s(N)$ are described by a recursive relation. For the Fibonacci sequence
$N=\{F_{1},F_{2},\ldots,F_{N-1},F_{N}\}$, H. Kubo and M. Goda 
\cite{HKubo} have investigated the statistical properties of 
$s(N)$ which are directly related to the characteristical exponents of 
self-similar wavefunctions. Taking  $(\psi_{0},\psi_{-1})=(-i,1)$, it 
is possible to show that $|\psi_{n}|^{2}=
{\displaystyle \gamma^{-2s(N)\times(-1)^{t(N)}}}$. The Landauer resistance 
of a finite chain N is given by $\rho_{N}=\frac{h}{e^{2}}\ \frac{R}{T}$ 
where T is the fraction of tunneling electrons transmitted from the 
system to the leads, and R is the reflected one. For a tight-binding 
model one relates the resistance to the total transfer matrix 
elements: $\rho_{N}=\frac{1}{4}(P^{2}_{N}(1,1)+P^{2}_{N}(1,2)
+P^{2}_{N}(2,1)+P^{2}_{N}(2,2)-2)$.
\begin{figure}[htbp]
\begin{center}
\leavevmode
\epsfxsize=5cm
\centerline{\epsffile{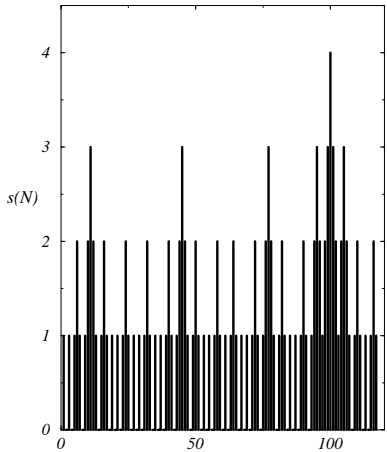}}
\narrowtext{\caption{Multifractal distribution of $|s(N)|$
 for the Fibonacci chain of 800 sites.}}
\end{center}
\label{sN}
\end{figure}
and one write the Fibonacci chain's resistance in a closed form
$$\bigl(\rho_{N}\bigr{)}_{\hbox{I}}
=\frac{1}{4}(\gamma^{2s(N)}+\gamma^{-2s(N)})-\frac{1}{2} 
=\biggl(\frac{\gamma^{s(N)}-\gamma^{-s(N)}}{ 2}\biggr{)}^{2}$$

The function $s(N)$ is illustrated on Fig.\ref{sN} which manifest 
self-similar pattern. When $s({N}_{0})=0$, transmission is 
perfect $T=1$. If one considers the Lyapunov exponents for finite length
 systems we get some estimates of the associated localization lengthes at
  a given energy \cite{Suto2}. By definition 
effective Lyapunov exponents (since localization length can be much larger than system size in the finite limit) for a chain of N sites are given by 
$\rho_{N}(E)=\frac{h}{2e^{2}} exp(\gamma_{N}(E)\times N)$ which is equivalent to $\gamma_{N}(E)=\frac{1}{N}|P_{N}(1,2)|^{2}$
 (given that $\rho_{N}(E)=|P_{N}(1,2)|^{2}$).

\begin{figure}[htbp]
\begin{center}
\leavevmode
\epsfxsize=5cm
\centerline{\epsffile{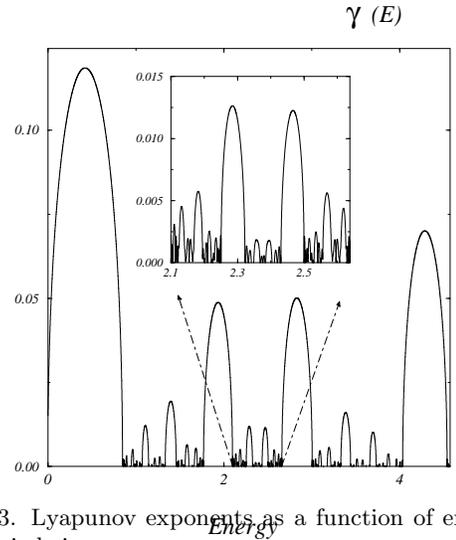}}
\narrowtext{\caption{Lyapunov exponents as a function of energy of a Fibonacci 
chain.}}
\end{center}
\label{Lyapu}
\end{figure}

For energies lying outside the spectrum are easily identified with stronger Lyapunov 
exponent as exemplified on Fig.\ref{Lyapu}. The cantor spectrum for 
eigenvalues is revealed by the corresponding self-similar pattern 
of localization lengthes distribution.

We now remind the effect of phason defect on the 
Landauer resistance obtained in \cite{RocheFas}. Taking the position of the 
defect, represented by a -BB- unit as an internal 
degree of freedom, at E=0, the properties of the 
matrices  ${\cal A,B, C}$ are shown to be independent of the position 
henceforth called $x_{P}$\cite{RocheFas}. By simple 
manipulations of the transfer matrix, we further demonstrate that the total transfer matrix associated to the chain with one phason and with  $N(i)$ links
is the same as the one without phason but with N(i+2) links. Accordingly, a generic form can be written down : 
$P_{N}$ (with either $a=d=0$, or $b=c=0$)

\beq
P_{N(i)}|_{I}=
\left(
\bea{cc}
 a & b \\
c & d \\
\ea
\right)
\eeq
then
\beq
P_{N(i)}|_{III}=
P_{N(i+2)}|_{I}=-
\left(
\bea{cc}
0  & -\gamma \\
1 & 0 \\
\ea
\right)
\left(
\bea{cc}
0 & -\frac{1}{\gamma} \\
1 & 0 \\
\ea
\right)
\left(
\bea{cc}
 a & b \\
c & d \\
\ea
\right)
\eeq

Landauer resistance was shown to be given by \cite{RocheFas}:

$$\bigl(\rho_{N}\bigr{)}_{\hbox{III}} 
=\Biggl(\frac{\gamma^{s(N)-1}-\gamma^{-s(N)+1}}{ 2}\Biggr{)}^{2}$$

By comparing the two expressions for the Landauer 
resistance one concludes that at $E=0$, the sign of
$\rho_{N}\bigr{)}_{\hbox{I}} -(\rho_{N}\bigr{)}_{\hbox{III}}$ is 
fluctuating as a function of chain-length, which means 
the phason defect does not alter the transport properties in the infinite limit.

We now consider the sequence constructed following 
the same algorithm but maximizing the number of phason defects as 
identified by ${\bf BB}$. We then calculated analytically the 
Landauer resistance $\rho_{N}|_{IV}$. For instance for chains with 
N=4 and N=17 links, one has :

\begin{eqnarray}
(B)&-&\hbox{ABBA}-(B)\nonumber\\
(B)&-&\hbox{ABBABBAABABBABBAA}-(B)\nonumber
\end{eqnarray}

With the same transfer matrices previously introduced
 ${\cal A, B}$ and  noticing that BAABAB and BAB are equal 
 respectively to ${\cal A}$ and ${\cal -B}=\tilde{\cal B}$, we show that the sequence of $P_{N}|_{IV}$ 
is generated by the following Fibonacci sequence $P_{N}={\cal A}, \ {\cal A}\tilde{\cal B}, \ {\cal A}\tilde{B}{\cal 
A}, {\cal A}\tilde{\cal B}{\cal AA} \tilde{\cal B}, {\cal A}\tilde{\cal B}{\cal AA}\tilde{\cal B}{\cal A}\tilde{\cal 
B}{\cal A}$, ${\cal A}\tilde{\cal B}{\cal AA}\tilde{\cal B}
{\cal A}\tilde{\cal 
B}{\cal AA}\tilde{\cal B}{\cal AA}\tilde{\cal B},...$. The analytical form of $\rho_{N}|_{IV}$ is shown to be given by

$$
\bigl(\rho_{N}\bigr{)}_{\hbox{IV}} 
=\Biggl(\frac{\gamma^{\tilde{s}(N)}-\gamma^{-\tilde{s}(N)}}{ 
2}\Biggr{)}^{2}$$

with $\tilde{s}(N)$ a new complex function of N calculated 
iteratively. The interesting point to observe is that the function
 $\rho_{N}|_{I}-\rho_{N}|_{III}$ for one phason defect changes its sign
  at each step $N\to N+1$,
 whereas $\rho_{N}|_{I}-\rho_{N}|_{IV}$ manifestes fluctuations on much 
 larger range. As an illustration, chains with number of sites 
 respectively equal to $N(i=1,2,3,4,5...17)=5-203$ sites follow 
$\rho_{N}|_{I}-\rho_{N}|_{IV}\geq 0$, whereas the behavior is opposite 
for chain from 330 to 456 sites, and so forth.

In conclusion, even for the highest density,
 such phason disruptions of quasiperiodic order are not able to 
break down the localization mechanism which do remain basically the 
same in the limit $N\to\infty$.

\section{Landauer Resistance of a Kr\"{o}nig Penney model with phasons}

Herafter we first rewrite the main steps of calculations as first described in
\cite{RocheFas}. We then perfom power-spectra calculations of  
Landauer resistance interference patterns to clarify the differences between 
localization properties and transmission abilities of critical states.
In the Kr\"{o}nig Penney model\cite{KP}, the potential describing the 
interaction of the electron with the lattice is represented by a sum 
of Dirac distributions  with intensity $V_{n}$ localized at $x_{n}$ 
$V(x)=\sum_{n} V_{n}\delta(x-x_{n})$, the $x_{n},V_{n}$  can be chosen as correlated variables, or 
uncorrelated. In between two successive scattering centers, the 
solution of the Schr\"{o}dinger equation is a linear combination of 
two plane waves: $
\Psi(x)=A_{n}e^{ik(x-x_{n})}+B_{n}e^{-ik(x-x_{n})}$ $(x_{n}\leq 
x\leq x_{n+1})$, the 1D wavevector $k>0$ is related to the energy $E$ through {\small 
$E={\hbar}^{2}k^{2}/2m$}. In the Kr\"{o}nig-Penney (KP) 
\cite{KP} model, a solution of the problem is constructed by 
imposing continuity conditions for the wavefunction and its derivative. 
For sake of simplicity, we choose the case where the intensity of 
scattering potential is constant =$\lambda$ , whereas scattering centers are 
quasiperiodically spaced  $\{({x}_{n}-{x}_{n-1}) \}= \{a,b \}= \{\tau,1\}$. The problem can then be 
written as followed :

\beq
\left(
\bea{c}
A_{n+1} \\ B_{n+1} \\ \ea\right)=\Lambda(n).
\left(
\bea{c} A_{n} \\ B_{n} \\ \ea\right)\nonumber
\eeq
with

\begin{center}
\beq
\Lambda(n)=
\left(
\bea{cc}
(1-{{i \lambda}\over{2 k}} ) e^{i k (x_{n+1}-x_n)} &
-{{i \lambda}\over{2 k}}  e^{i k (x_{n+1}-x_n)} \\
\ & \  \\
{{i \lambda}\over{2 k}} e^{-i k (x_{n+1}-x_n)} &
(1+{{i \lambda}\over{2 k}} ) e^{-i k (x_{n+1}-x_n)}\\
\ea
\right)\nonumber
\eeq
\end{center}

within this framework, Landauer resistance is given by
$\rho_{N}= |P_{n}(1,2)|^2 \ \ \ \hbox{with} \ \ \ P_{n}= 
\Lambda(F_n) ...\Lambda(1)$. Furthermore, a renormalization group associated with 
the Fibonacci chain enables to map the electronic spectrum 
to a dynamical system associated with the traces of the transfer 
matrices. The following energy-dependent invariant of the Kr\"{o}nig-Penney
model is convenient to consider \cite{Invar}. In our case, we find  
$\hbox{I(k)}=\frac{\lambda^2 \sin^2 k(a-b)}{4 {k}^2}$ whose zeros
 are given by ${k_{s}}=n\pi/(a-b)$, with n integer. These points 
 refered as conducting points are interesting since they 
correspond to the commutation of the transfer matrices,
$[ P_{F_{n}},P_{F_{n+1}} ]=0 \ \ \hbox{given that} \ \ 4 
\hbox{I}+2=\hbox{Tr}(P_{F_{n}}.P_{F_{n+1}}.P_{F_{n}}^{-1}.P_{F_{n+1}}^{-1})$.
 Besides elementary matrices also commute :
\endmulticols\widetext
\vspace{-6mm}\noindent\underline{\hspace{87mm}}
\beq
\left[ \Lambda (\Delta x=a),\Lambda (\Delta x=b) \right ] =
\lambda \  {{e^{i k (a-b)}}\over {4 k^2}}\; (1-e^{2 i k (b-a)}) 
\left(
\bea{cc}
\lambda & \lambda-2 i k \\
-\lambda-2 i k& -\lambda\\
\ea
\right)\nonumber
\eeq
\noindent\hspace{92mm}\underline{\hspace{87mm}}\vspace{-1mm}
\multicols{2}
\noindent
The resistance can be written down analytically at ${k_{s}}$
\beq
\rho_{N}|_{_{_{k=k_s}}}=\biggl(\frac{\lambda}{2 k_s} \biggr)^2
\ \ \frac{\sin^2 N \varphi}{\sin^2 \varphi}\eeq
with $\varphi$, depending on $k^{2}_{s}$ and $\lambda$, 
and is defined equivalently by 
$\cos \varphi=\cos k_s a+ \lambda/{2 k_s} \sin k_s a$ or $\cos \varphi=\cos k_s b+ \lambda/{2 k_s} \sin k_s b$. The 
anaytical form of {\small 
$\rho_{N}|_{_{_{k=k_s}}}$} indicates that when $N\to\infty$, the 
Landauer resistante oscillates but remains bounded, so that the 
energies $k^{2}_{s}$ correspond to states that lead to best 
transmission, reminding that localized states will display exponantial 
increase of resistance. Following the discussion of \cite{Suto,Baake}, 
if ones restricts the study for energies 
$k^{2}=(k_{s}+\varepsilon)^{2}$ in the vicinity of $k_{s}$ where 
highest density of eigenvalues are expected in the infinite limit
(hereafter $k_{s}=\pi/(\tau-1), (a=\tau, b=1)$ is taken 
without loss of generality). Two transfer matrices are defined
\endmulticols
\vspace{-6mm}\noindent\underline{\hspace{87mm}}
\beq
{\Lambda}_{A}=
\left( 
\bea{cc}
(1-\frac{i\lambda}{2k}){e}^{ik\tau} & - \frac{i\lambda}{2k} 
{e}^{ik\tau} \\
\ & \ \\
\frac{i\lambda}{2k} {e}^{-ik\tau} & 
(1+\frac{i\lambda}{2k}){e}^{-ik\tau}\\
\ea
\right) 
\ \ \ \ 
\Lambda_{B}=
\left( 
\bea{cc}
(1-\frac{i\lambda}{2k})e^{ik} & - \frac{i\lambda}{2k} {e}^{ik} \\
\ & \ \\
\frac{i\lambda}{2k} {e}^{-ik} & (1+\frac{i\lambda}{2k}) {e}^{-ik}\\
\ea
\right)
\eeq
\noindent\hspace{92mm}\underline{\hspace{87mm}}\vspace{-2mm}
\multicols{2}
\noindent
Choosing $\lambda$ so that {\small 
$\rho_{N}|_{_{_{k=k_s}}}=0$} exactly at $k={k_s}$, which lead to $N-1$ 
values for $\lambda$, given by 
${\lambda}_{s}= {{2 k_s (\cos \varphi_s-\cos k_s )}\over {\sin k_s}}$
with the phase $\varphi_s$ defined through {\small ${\varphi}_{s}=(m 
\pi)/ N$} with m=1,...,N-1. The N-1 values $\varphi_s$ are 
symetrically distributed around  $\pi/2$ and cover densely the 
half-upper trigonometric plane when increasing N \cite{RocheFas}. 
Reminding that $x_{P}$ is taken as 
the position of phason defect ({\bf BB}) in the chain, with 
$x_{P}\in [1,P]$ an integer and $P$ the maximum number of 
position for a given chain, we know recall main important patterns \cite{RocheFas}
and perform power spectra of $\rho_{N}(x_{P},\varphi_s,\varepsilon)$.

\begin{figure}[htbp]
\begin{center}
\leavevmode
\epsfxsize=6cm
\centerline{\epsffile{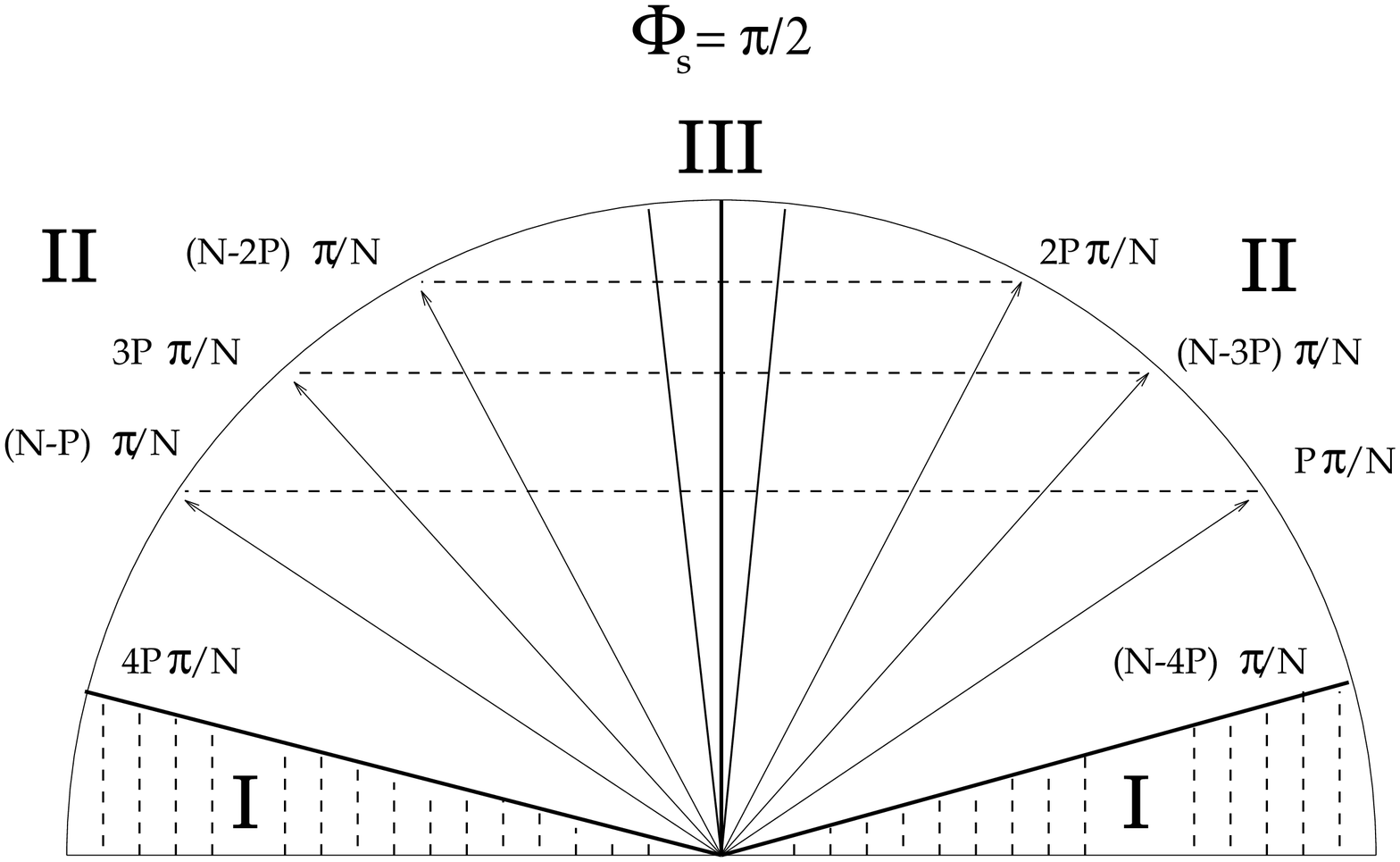}}
\narrowtext{\caption{Energy dependent phase diagram.}}
\end{center}
\label{figdf}
\end{figure}

\begin{figure}[htbp]
\begin{center}
\leavevmode
\epsfxsize=6cm
\centerline{\epsffile{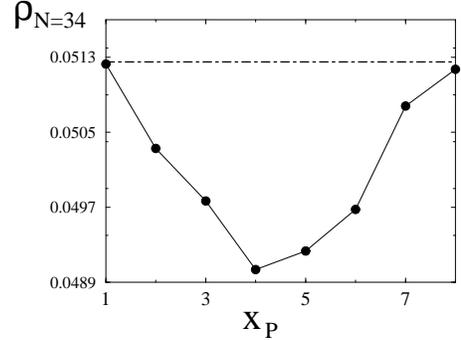}}
\narrowtext{\caption{Landauer resistance as a function of phason position
 $x_{P}$, for $\varepsilon=10^{-4}$, $N=35$ sites and m=1.}}
\end{center}
\label{figR34}
\end{figure}
\begin{figure}[htbp]
\begin{center}
\leavevmode
\epsfxsize=6cm
\centerline{\epsffile{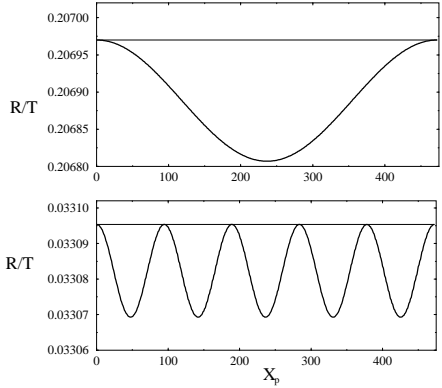}}
\narrowtext{\caption{Landauer resistances as a function of phason position 
${x}_{P}$ (N=2001,P=472) for $\varepsilon=[10^{-9},10^{-6}]$ 
and m=5. For higher values of $\varepsilon$, then $\rho_{N} \to \infty$ 
which indicates that energy lies within a gap, or may be associated 
with a localized state.}}
\end{center}
\label{fig1-5}
\end{figure}

%
Numerical accuracy is checked through $P_{N}(1,1)^{2}-1=P_{N}(1,2)^{2}$ which
 gives the resolution. The Landauer resistance calculated in $k=k_{s}$ is always found 
 to be $\sim 10^{-12}-10^{-13}$ and gives thus the numerical uncertainty 
 on $\rho_{N}$. We now present the main results as sketched on the phase diagram in 
Fig.\ref{figdf}. According to the variable $\varphi_{s}$ ones identifies several features. First, note that amazing pseudo-symmetry 
is found around $\pi/2$ for $\rho_{N}(x_{P},m,\varepsilon)$ which do 
not correspond to any symmetry in the scattering potential (see below Fig.\ref{figR34}-\ref{fig1-5} for illustration).

In the zones refered as zone-{\bf I} 
the phason reduces the resistance for 
energies sufficiently close to the conducting points (see 
Fig.\ref{figR34}-\ref{fig1-5}). Two symmetrical zones are 
$ m <  N-4P, \ \hbox{and} \ m  >  4P$
with ${\rho_{N}}\mid_{\hbox{I}}(x_{P},m,\varepsilon)>
{\rho_{N}}\mid_{III}(x_{P},m,\varepsilon)$. In other words, for that 
values of parameters the Fibonacci chain becomes more conductive upon 
introduction of local phason. This effect is a pure result of quantum 
intereference at low temperature and has been shown for tunneling 
energies close to the ones of conducting points, so for the electrons 
that can better contribute to conduction mechanism. We illustrate 
these behaviors first on Fig.\ref{figR34} which allow to write down 
all the 8 different inequivalent defected-chains as described below for 
clarity for a system with 35 sites:
\endmulticols\widetext
\vspace{-6mm}\noindent\underline{\hspace{87mm}}
\begin{center}
\beqn
FIBO&:& --B--ABAABABAABAABABAABABAABAABABAABAAB--B--\nonumber\\
Def1&:& --B--A{\bf BB}ABAABAABABAABABAABAABABAABAABAB--B--\nonumber\\
Def2&:& --B--ABAABA{\bf BB}AABABAABABAABAABABAABAABAB--B--\nonumber\\
Def3&:& --B--ABAABABAA{\bf BB}ABAABABAABAABABAABAABAB--B--\nonumber\\
Def4&:& --B--ABAABABAABAABA{\bf BB}ABAABAABABAABAABAB--B--\nonumber\\
Def5&:& --B--ABAABABAABAABABAABA{\bf BB}AABABAABAABAB--B--\nonumber\\
Def6&:& --B--ABAABABAABAABABAABABAA{\bf BB}ABAABAABAB--B--\nonumber\\
Def7&: &--B--ABAABABAABAABABAABABAABAABA{\bf BB}AABAB--B--\nonumber\\
Def8&:& --B--ABAABABAABAABABAABABAABAABABAA{\bf BB}AB--B--\nonumber
\eeqn
\end{center}
\noindent\hspace{92mm}\underline{\hspace{87mm}}\vspace{-3mm}
\multicols{2}
\noindent
Besides, the function ${\rho_{N}}\mid_{\hbox{III}}(k,m,x_{P})$ 
given a kind of interference pattern described by the position of 
phason defect ${\rho_{N}}\mid_{\hbox{III}}(k,m,x_{P})\sim \alpha(m/P)\  
\sin\bigl( \frac{2m\pi}{P} x_{P}\bigr)$. This is further represented 
on Fig.\ref{fig1-5} for chains with N=2000 links, and m=1,5. The 
curves for m=1 (Fig.\ref{figR34} and Fig.\ref{fig1-5}(top)) and different chain lengths shows that the 
interference pattern encloses a memory of self-similarity.

\begin{figure}[htb]
\begin{center}
\leavevmode
\epsfxsize=6cm
\centerline{\epsffile{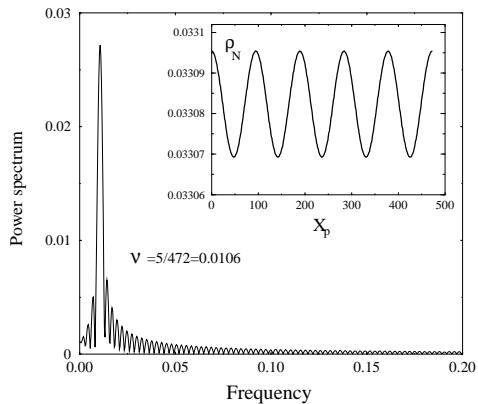}}
\narrowtext{\caption{Power spectrum ofthe pattern given in the inset for m=5 and 
same parameters as previous figure (bottom).}}
\end{center}
\label{fig6}
\end{figure}

The power spectrum of m=5 curve is given on Fig.\ref{fig6}, where 
the only eigenfrequency agrees with $m/P=5/472=0.0106$ in this case. 
Superimposed small oscillations are an unphysical effect due to a 
Fourier transform of a finite signal.

\begin{figure}[htbp]
\begin{center}
\leavevmode
\epsfxsize=6cm
\centerline{\epsffile{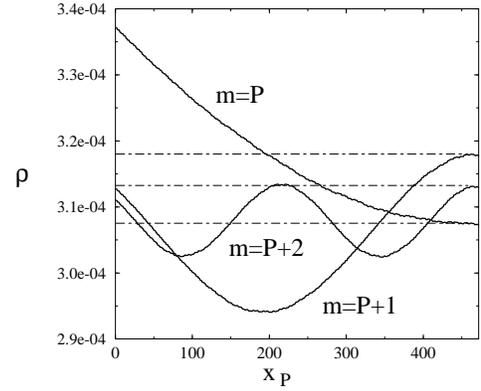}}
\narrowtext{\caption{Regular evolution of interference pattern for $m=P, P+1,P+2$ 
and $N=2000$, $P=472$. For $\varphi_{s}=P\pi/N$, Fibonacci chain is 
always less resistive that the imperfected one whereas transition occurs 
for $m=P+1,P+2,...$. dashed curves are the values for Fibonacci chains 
with same parameter m as imperfected one.}}
\end{center}
\label{figr3}
\end{figure}

Let's move forward to zone-{\bf II}. for $N-4P<m<4P$
situations becomes more complex but recurrent simple 
($m-P$)-periodic oscillatory patterns are found around the values $\varphi_{s}= 
\{ (N-4P)/N , \ (N-3P)/N, \ (N-2P)/N , \ (N-P)/N \} \ \pi$ and 
symetrically for
$\tilde{\varphi}_{s}= \{ 4P/N , \ 3P/N , \ 2P/N , \ P/N \}  \pi$. In 
these regions of parameter space, there is an genuine transition from 
systematic increasing to decreasing (and vice versa) of the electronic resistance upon
introduction of phason, as exemplified on Fig.\ref{figr3}.

\begin{figure}[htbp]
\begin{center}
\leavevmode
\epsfxsize=7cm
\centerline{\epsffile{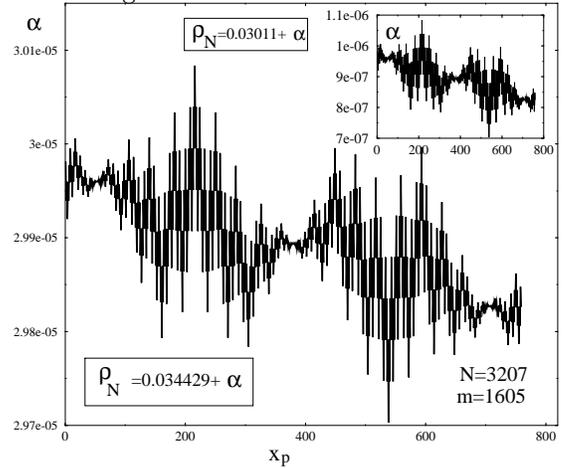}}
\narrowtext{\caption{Landauer resistance for values of scattering potential which 
unravel self-similar patterns as described in the text.}}
\end{center}
\label{figrc1}
\end{figure}

As far as zone-{\bf III} is concerned, and related to some interval 
around  $\varphi\sim\frac{\pi}{2}$, self similar patterns are 
observed which suggest that ${\rho_{N}}\mid_{\hbox{III}}(x_{P})$ 
reveal critical states which are robust against phason disorder as 
found in the TBM for E=0 case. The typical patterns represented on 
Fig.\ref{figc1} actually encloses oscillations of resistance which 
smaller oscillations are described by some coefficients $s(n)$ as 
described previously. 

\begin{figure}[htbp]
\begin{center}
\leavevmode
\epsfxsize=7cm
\centerline{\epsffile{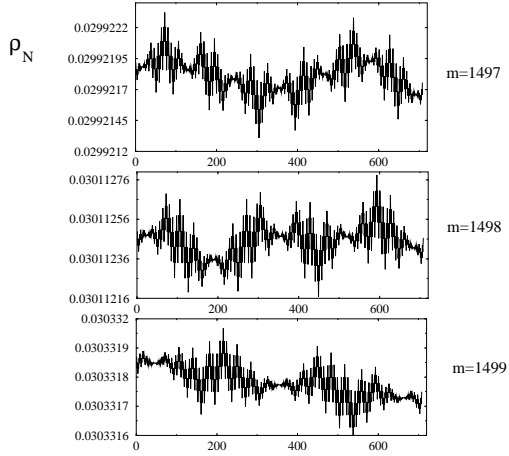}}
\narrowtext{\caption{Interferences pattern for several values of scattering 
potentials close to $\varphi\sim\frac{\pi}{2}$.}}
\end{center}
\label{figc3}
\end{figure}
\noindent
On the Fig.\ref{figc3}, the case with 3000 sites is considered and 
differents values of the scattering potential in the vicinity of the 
symmetry point $\varphi=\frac{\pi}{2}$. For m=1497, the potential is
$\lambda \sim 3.918$, for m=1498 $\lambda \sim 3.929$ and 3.941 for 
m=1499. Patterns exhibit small differences, and their Fourier spectrum 
is actually very similar (see below).

\begin{figure}[htbp]
\leavevmode
\epsfxsize=8cm
\centerline{\epsffile{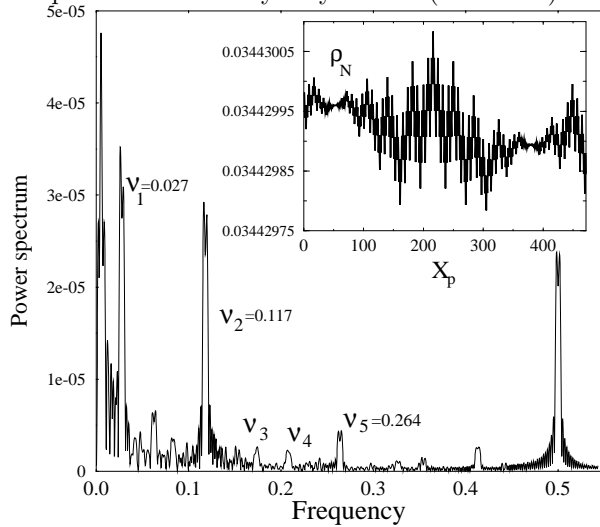}}
\narrowtext{\caption{Self similar interference pattern for N=3207, P=757 and J=1605. 
Stability is ensured from $\varepsilon=10^{-3}$ down to 
the numerical resolution limit.}}
\label{figc1}
\end{figure}

\begin{figure}[htbp]
\begin{center}
\leavevmode
\epsfxsize=6cm
\centerline{\epsffile{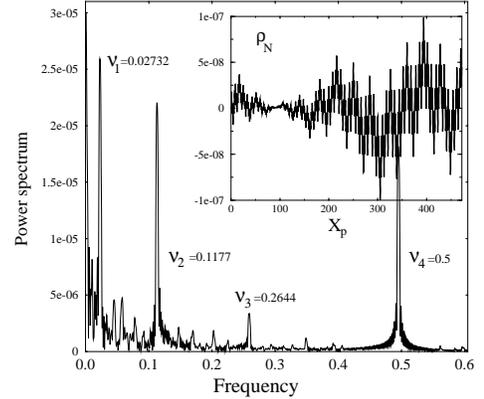}}
\narrowtext{\caption{Same as previous figure with different m.}}
\end{center}
\label{figc2}
\end{figure}

\noindent
The power spectra of different similar patterns are 
given on Fig.\ref{figc1} and Fig.\ref{figc2} which show that 
superimposed frequencies are identical. The highest frequency is given 
by $\nu=0.5$ which is related to the change of $\rho_{N}(x_{P}\to 
x_{P}+1)$. On 
respective figures five unambiguous frequencies have been located and 
named $\nu_{n=1,5}$. This prove that all these self-similar
patterns are actually described by exactly the same function which is 
a superposition of several independent frequencies convoluate with a 
function defined by a serie of $s(N)$-type coefficients.
Concerning all the discussed zones-{\bf I, II, III},
simple forms for energies sufficiently close to conducting points. 
 Then $\rho_{N}(x_{P},\varepsilon)$ result from a superposition of one or several frequencies. For larger energies
  one finds fluctuations as those analyzed in the inset of Fig.\ref{figc3}. As 
  energies get farther to conducting points ${k}_{s}=\pi/(\tau-1)$ the 
  resistance is sharply increasing on several order of magnitudes. The 
  power spectrum in this case reveal much more eigenfrequencies, with 
  the interesting emergence of frequencies with similar amplitudes 
  refered as $\nu$ and $\omega$. In all these cases the respective 
  behaviors of the Landauer resistance of the perfect Fibonacci chain versus the 
  imperfect one follows complicated random fluctuations. 
 
The different behaviors found in these studies suggest that in some case 
local disruption of long range quasiperiodic order has improved the
 conductive ability of the chain in a systematic manner. Analyzing the interference pattern of the Landauer resistance as a function of 
  phason defect suggest that extendedness (as a localization 
  properties of available states at such energies) has also been 
  jointly improved. This is shown by a bounded and simple oscillatory 
  pattern for the resistance, common to what is found for extended 
  eigenstates in a periodic systems.

\begin{figure}[htbp]
\begin{center}
\leavevmode
\epsfxsize=6cm
\centerline{\epsffile{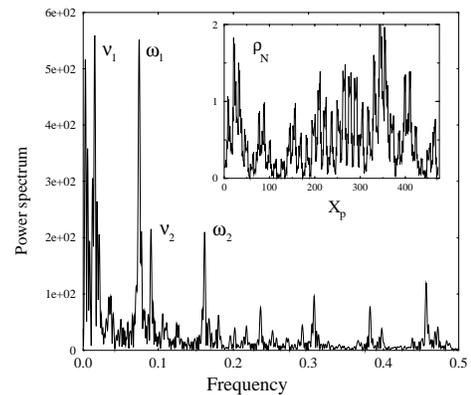}}
\narrowtext{\caption{Landauer resistance interference pattern for 
$\varepsilon=0.5$, N=2000 and P=472.}}
\end{center}
\label{fig3}
\end{figure}

To conclude this study, one stresses that first conducting points 
indeed seem to be the location of high density of eigenenergies as 
discussed in \cite{Suto,Baake} and Kr\"{o}nig Penney model
 has been able to unravel specific quantum interference effects of
 phason disorder on localization and propagation properties, hidden 
 in the treatment of TBM, and appealing in the context of 
 quasicrystalline materials. Here following earlier results
 \cite{RocheFas}, we have found that multi phason defects 
 in TBM at E=0, do not alter the transmission abilities of 
 corresponding states, and that interference patterns as revealed by
 $\rho_{N}(x_{P},\varepsilon)$ and their power-spectra analysis is an 
 original way discovering how phason defects affect jointly
 localization and transport modes.

\section{Quantum interference mechanisms in high-dimensionnal 
quasicrystals}

Some effort to investigate quantum interferences effects in small
quasiperiodic penrose approximants have been made \cite{Moulopoulos}. 
Here we propose how quantum interferences on 
both sides of a metal-insulator transition in real materials might be 
analyzed. Indeed, weak localization regime has been found in experiments for some 
quasicrystalline materials (AlCuFe,\ldots) whereas other systems such as 
AlPdRe-quasicrystals behave differently being very close 
to a metal-insulator transition\cite{Berger}. Two different focus may be considered for a general understanding of 
quantum interferences in quasicrystals. First, as there exist 
approximant phases (periodic) sharing the same behavior, weak 
localization correction beyond Drude approximation should apply for 
that systems as well as for corresponding quasicrystals. The 
only recurrent approximation when solving the cooperon diffusion equation 
is to assume that scatters are uncorrelated i.e
$\<{\cal U}(r){\cal U}(r')\>_{disorder}=cu^{2}\delta(r-r')$
(${\cal U}(r)= \sum_{i=1}^{N}\  {\cal U}(r-R_{i})$, c the impurity 
concentration, u typical strength). The calculation of 
the quantum correction of the conductivity in this regime is
 enclosed in phase factor interferences of the two-particle 
 Green's function  $\<GG^{*}\>$ . By performing configuration averaging beyond the mean free path legnth
scale, then $\< GG^{*} \>_{disorder}$ reduces phase interference to 
the ($k'=-k$)-Cooperon pole, as 
a consequence of time reversal symmetry, the possibility to have a 
coherent distribution of scatters is usually neglected.

However, assuming that the distribution of scatters is
constrained to, let's say for sake of illustration, a mirror-plane symmetry, i.e 
$\forall \alpha \in  \{ R_{\alpha}, \alpha=1,N\}$ there is a site
 $R_{\beta}$ such that $R_{\beta}=-R_{\alpha}$, then without performing any 
 diagrammatic expansion, we just notice that weak localization related with average of the potential 
 scattering $\< {\cal U}(r){\cal U}(r'){\>}_{disorder}$ corresponding to
 phase factors 
 
 $$\sum_{\alpha\beta}\< {e}^{-i(k\ R_{\alpha}+p\ 
R_{\beta})}{\>}_{disorder} =\frac{1}{\Omega^{2}}\ \int d^{3}R_{\alpha}d^{3}R_{\beta} {e}^{-ikR_{\alpha}}
\ {e}^{-ipR_{\beta}}$$

\noindent
 will display new terms associated to above-mentionned symmetry, 
$\< {e}^{-i(k\ R_{\alpha}+p\ R_{\beta})} {\>}_{disorder}=
 \< {e}^{i(p-k) R_{\alpha}} {\>}_{disorder}=\delta_{p-k,0}$ that will 
 increase the contribution of phase interferences.  
 Say in another way, if a double symmetrical loop crosses in the mirror plane and in a region with extension less or equal to $\lambda_{F}$, then four equivalent pathes 
will interfere at the returning point instead of the usual two of the weak localization scheme, resulting in a total interference
  amplitude will be 4 times stronger ($\mid {\cal A}_{I}+{\cal A}_{II}+{\cal A}_{III}+{\cal A}_{IV}\mid 
  ^{2}=16\mid {\cal A}\mid ^{2} $). Similar
   ideas have been already introduced in context
    of mesoscopic physics \cite{Baranger} but we propose here that 
    they may serve as a path to follow a metal-insulator
     in quasicrystals in which even disorder (such as phasons) may keep
      some strong correlated properties\cite{Roche-un}.

The second focus is suggested by the close proximity of a metal-insulator 
transition and gives a very dissimilar weight to quantum interferences. Any critical point of an electronic localisation-delocalisation transition
can essentially be described by its anomalous diffusion which means that two-particle Green function reads
 $\< |G^{+}(r,r';E)|^{2}\>\sim|r-r'|^{-\eta +2-D}$ with $\eta$ a critical 
 exponent (and the propagator representes the transition 
 probability in real space of an electron of E energy from site
$|r\>$ to $|r'\>$), and which directly affect the conductivity of the
system since 

\beq
\sigma_{DC}=\frac{2e^{2}}{h}\lim_{\varepsilon\to 0} 
4\varepsilon\int d^{D}r r^{2}\< |G^{+}(r,r';E)|^{2}\>_{disorder}
\label{eqs}
\eeq

\noindent
In the case where all states are localized (insulating side), 
the average is carried out over random configurations of frozen 
disorder and given that $\<|G^{+}(r,r';E)|^{2}\>_{imp}\sim \exp^{-|r-r'|/\xi_{E}}$ ones easily 
recovers that $\sigma_{DC}\to 0$ at zero temperature. At the critical 
point, the power law describing electronic propagation has been
 illustrated in quantum Hall regime (D=2) for 
 length scales in the regime of multifractality, i.e $l\ll r\ll\xi$, 
 $\eta\simeq 0.38\pm 0.04$\cite{QHE} or analytically in 
 hierarchical potential \cite{Jona} and quasiperiodic systems\cite{Bell}
  and numerically in 3D quasiperiodic systems\cite{RocheJMP,RochePRL1}. No characteristic length scales, such as 
the mean free path, can be defined and interference effects are 
intrinsically defining the dominant transport mechanism. A simple 
analysis of the equation 12 shows that according to the strength
of criticality, as manifested by the value of the exponent, conductivity 
can be zero or infinite at zero temperature for a perfect system. 
Elastic scattering allow to work out the exponents of the 
anomalous Drude formula\cite{Bell,RochePRL1} at very low temperature.

 At finite temperature, a dissipation mechanism results in a cut-off in the
integral, and one always gets a finite value for the resistance. Since the interesting
point occurs when multifractality dominate the transport regime, it 
is then important to estimate temperature dependent transport. In 
quasicrystals, critical states are associated with the so-called 
anomalous Drude law ($\sigma\sim \tau^{2\beta-1}$) \cite{RocheJMP,Bell}, where $\tau$ is supposed to be the relevant
 inelastic scattering rates at a given temperature. From a
scaling analysis of the Kubo formula on periodic quasicrystalline 
approximants, and from the knowledge of exact eigenstates a scaling behavior was revealed
${\displaystyle
 \Delta\sigma(T)\sim T^{\eta},\ \ \hbox{\small with}\ \  \eta\simeq 1.25}$
in good agreement with the values obtained
experimentally for several quasicrystalline and approximant
phases at temperature $\geq 10K$ \cite{Gignoux-I,Kimura-I,Lalla-I}.

In both side of the metal-insulator transition, the presence of 
phasons as demonstrated in 1D systems may weeken the interference
 effects while destroying quasiperiodic 
long range order, resulting thus in a unconventional mechanism. 
Working this out analytically in realistic models remains a great 
challenge.

\section{Conclusion}

Several features occurring in the electronic
transport of 1D quasicrystals have been reviewed. Quantum interferences 
have been shown to be interestingly altered by phason 
defects which may be a natural disruption of quasicrystalline disorder. 
In quasicrystals QIE may be sustained by 
different mechanisms as discussed in the second part. Revealed pattern 
in 1D-systems may keep some generality in higher dimension. 

\acknowledgments 
S.R. is indebted to the European Commission for financial support (Contract
ERBIC17CT980059), and to Prof. T.  Fujiwara from Department of Applied Physics of
Tokyo University for hospitality.  Kostas 
Moulopoulos is deeply acknowledged for many enlightening discussions 
and encouragements.

\vfill\eject
Figure captions

Fig1
Chain $N=4$: ABAA connected with perfect leads.

Fig2
Multifractal distribution of $|s(N)|$ for the Fibonacci chain of 800 sites.

Fig3
Lyapunov exponents as a function of energy of a Fibonacci chain.

Fig4
Energy dependent phase diagram.

Fig5
Landauer resistance as a function of phason position 
$x_{P}$, for $\varepsilon=10^{-4}$, $N=35$ sites and m=1.

Fig6
Landauer resistances as a function of phason position 
${x}_{P}$ (N=2001,P=472) for $\varepsilon=[10^{-9},10^{-6}]$ 
and m=5. For higher values of $\varepsilon$, then $\rho_{N} \to \infty$ 
which indicates that energy lies within a gap, or may be associated 
with a localized state.

Fig7
Power spectrum ofthe pattern given in the inset for m=5 and 
same parameters as previous figure (bottom).

Fig8
Regular evolution of interference pattern for $m=P, P+1,P+2$ 
and $N=2000$, $P=472$. For $\varphi_{s}=P\pi/N$, Fibonacci chain is 
always less resistive that the imperfected one whereas transition occurs 
for $m=P+1,P+2,...$. dashed curves are the values for Fibonacci chains 
with same parameter m as imperfected one.

Fig9
Landauer resistance for values of scattering potential which 
unravel self-similar patterns as described in the text.

Fig10
Interferences pattern for several values of scattering 
potentials close to $\varphi\sim\frac{\pi}{2}$.

Fig11
Self similar interference pattern for N=3207, P=757 and m=1605. 
Stability is ensured from $\varepsilon=10^{-3}$ down to 
the numerical resolution limit.

Fig12
Same as previous figure with different m.

Fig13.
Landauer resistance interference pattern for 
$\varepsilon=0.5$, N=2000 and P=472.

\end{document}